\documentclass[aps,prl,reprint,superscriptaddress, amsmath,amssymb]{revtex4-1}

\usepackage{graphicx}
\usepackage{dcolumn}
\usepackage{bm}

\begin{document}


\title{Anomalous dynamic behaviour of optically trapped high aspect ratio nanowires}


\author{Wen Jun Toe}
\affiliation{School of Physics, The University of New South Wales, Sydney NSW 2052, Australia}

\author{Ignacio O. Piwonka}
\affiliation{School of Mathematics and Statistics, The University of New South Wales, Sydney NSW 2052, Australia}

\author{Christopher Angstmann}
\affiliation{School of Mathematics and Statistics, The University of New South Wales, Sydney NSW 2052, Australia}

\author{Qiang Gao}
\author{Hark Hoe Tan}
\affiliation{Research School of Physics and Engineering, The Australian National University, Canberra ACT 2601, Australia}

\author{Chennupati Jagadish}
\affiliation{Research School of Physics and Engineering, The Australian National University, Canberra ACT 2601, Australia}

\author{Bruce Henry}
\affiliation{School of Mathematics and Statistics, The University of New South Wales, Sydney NSW 2052, Australia}

\author{Peter J. Reece}
\email{p.reece@unsw.edu.au}
\affiliation{School of Physics, The University of New South Wales, Sydney NSW 2052, Australia}



\date{\today}

\begin{abstract}
We investigate the dynamics of high aspect ratio nanowires trapped axially in a single gradient force optical tweezers. A power spectrum analysis of the Brownian dynamics reveals a broad spectral resonance of the order of a kHz with peak properties that are strongly dependent on the input trapping power. Modelling of the dynamical equations of motion of the trapped nanowire that incorporate non-conservative effects through asymmetric coupling between translational and rotational degrees of freedom provides excellent agreement with the experimental observations. An associated observation of persistent cyclical motion around the equilibrium trapping position using winding analysis provides further evidence for the influence of non-conservative forces. 
\end{abstract}


\maketitle

Optical trapping studies of low-symmetry objects, such as rods, ellipsoids and oblate spheroids, provides new dimensions to the study of optical trapping dynamics. In addition to more complex hydrodynamic interactions, non-spherical objects are subject to preferential trapping orientations and associated restoring torques when displaced away from equilibrium. When the trapping point is positioned arbitrarily with respect to the geometric centre, perturbative forces can lead to complex cross-coupling between translational and rotational motion\cite{Mihiretie:2014aa}. Shape mediated effects have been discussed in the context of  tailored optical force profiles, including the generation of optical lift \cite{Swartzlander:2011aa}, negative radiation pressure \cite{PhysRevLett.107.203602}, and have consequently been implemented in a practical setting for passive force clamping in photonic force microscopy \cite{B.:2014aa}. 

One interesting peculiarity that has seen growing attention is the tendency of low-symmetry particles to undergo persistent oscillatory motion in thermal equilibrium. This type of motion was first articulated for optically trapped rods by Simpson and Hanna, and is related to the influence of non-conservative optical scattering forces, which induces coupling between rotation and translation \cite{PhysRevE.82.031141}. For rods aligned nominally along the trapping axis (z-axis), the predicted motion is to circulate around a central trapping point within a plane (e.g XZ plane), coupled with a rocking of the rod orientation around the orthogonal axis ($\theta_{y}$). The concept has its genesis in an earlier study by Roichman \emph{et al.} who explored the influence of the non-conservative scattering force in producing circulating currents in the stochastic trajectories of spherical particles in an optical tweezers \cite{PhysRevLett.101.128301}. These concepts have been extended by  Saberi and Gittes who consider  general non-conservative forcing effects in optical tweezers \cite{Saberi:11}.

\begin{figure}[htb]
\centerline{\includegraphics[width=0.75\linewidth]{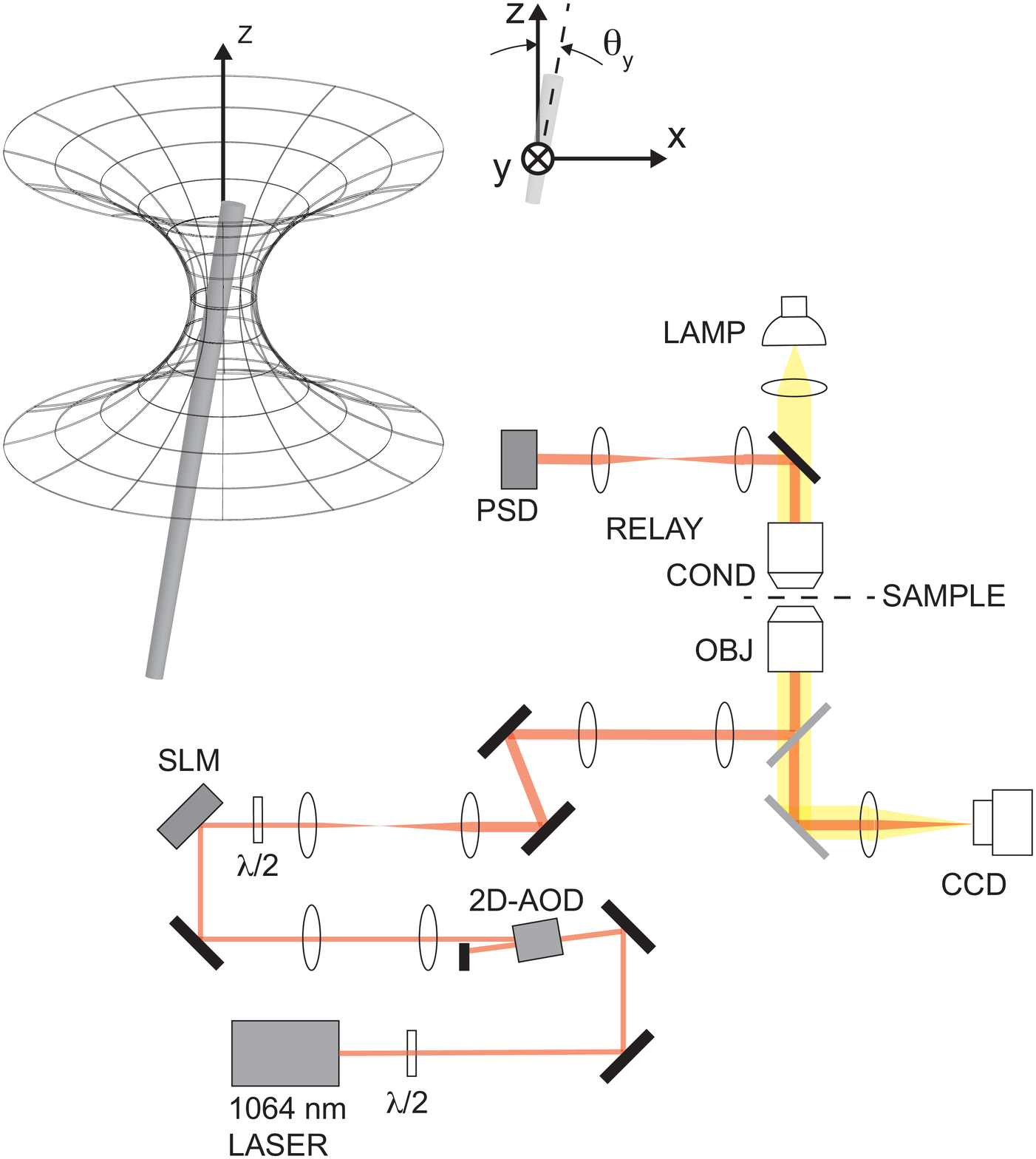}}
\caption{\label{fig:apsTOEfig1}(Color online) Schematic of our experimental setup and geometry within the optical trap with relevant coordinates and angles. In the experimental arrangement a diode pumped solid state laser is used with adaptive optics to create a diffraction limited spot within the focal plane of the objective (OBJ). The nanowires are trapped near the end of the wire and aligned along the laser axis. Light scattering from the nanowire is collected by the condenser (COND) and used to perform back focal plane interferometry with a position sensitive detector (PSD)}
\end{figure}

Our interests lie in experimental investigations of weakly tapered, high refractive index, indium phosphide (InP) nanowires of uniform composition, held in a linearly polarised, gradient force optical tweezers \cite{doi:10.1021/nl200720m}. Nanowires are a particular subclass of non spherical objects with nanoscale radial dimensions and lengths that extend to several micrometers, giving aspect ratios in excess of 100:1. For trapping experiments, the extended axial dimension can exceed the depth of focus of the trapping beams by several times the Rayleigh range, thereby creating a trapping force that is localised to a particular point on the axis. Earlier studies have shown that the equilibrium trapping position of these nanowires has been determined to be near one end, with the major segment of the length sitting below the trap far from the geometric centre \cite{doi:10.1021/nl304607v}, which is qualitatively  consistent with modelling of tapered rods \cite{Simpson201391}. The optically trapped nanowires are aligned closely to the optical axis of the tweezers and experience small perturbations from the equilibrium point due to stochastic motion. Note that these trapping conditions exclude the possibility of major orientation changes that are observed in other studies \cite{Neves:2010aa, Tong:2010aa}

Here we present experimental evidence for the presence of resonant fluctuations in the stochastic motion of optically trapped nanowires. A power spectral density analysis of the nanowire motion reveals a broad resonance peak of the of the order of a kHz, which is strongly dependent on trapping power. Modelling of the dynamical equations of motion of the trapped nanowire with asymmetric cross coupling terms between translational and rotational degrees of freedom provide good agreement with the experimental observations, suggesting that the effect is due to the action of non-conservative forces. Further, a winding number analysis of the Brownian reveals clear cyclical motion around the trapping point, which is another characteristic property of non-conservative effects.

Our experimental apparatus consists of a single beam gradient force optical trap generated at the focus of a high numerical aperture oil immersion microscope objective lens (Nikon E Plan Achromat 100x 1.25NA) using a Nd:YAG laser emitting at 1064nm (Laser Quantum IR Ventus). Dynamic steering of the laser focus is provided by a two-axis acousto-optic deflector (AOD) (Gooch and Housego 45035 AOBD) and spatial light modulator (SLM) (Hamamatsu LCOS-SLM x10468-03) positioned within the optical train at planes conjugate with the back aperture of the objective. SLM-based aberration correction and overfilling of the back aperture are used to produce a diffraction limited spot optimised for optical trapping. The power of the trapping laser at the imaging plane is controlled by the AOD and set between 10 mW and 90 mW. A schematic diagram of the trapping arrangement is shown in FIG~\ref{fig:apsTOEfig1}.

The indium phosphide (InP) nanowires used in the study are epitaxially grown on InP (111)B substrates by metal-organic chemical vapour deposition (MOCVD) using gold nanoparticle catalysts. The nanowires are between 5 and 10$\mu$m have a tapered aspect of approximately 0.3$^{\circ}$ as characterized by electron microscopy. Further information on the growth process can be found elsewhere, \cite{Paiman:2009aa}.

For tweezers measurements, the nanowires are dispersed into solution by sonication and transferred to a hermetically sealed chamber at a dilution where single nanowires can be isolated. Visualising the nanowires in the trap is achieved using a charge-coupled device (CCD) camera (AVT Stingray) together with diascopic illumination supplied by the condenser. Back focal plane interferometry is performed by collecting the forward scattered light from the trapped nanowires with a high numerical aperture condenser lens (Olympus S Plan Fluor 40x 0.6NA) and projected onto a position sensitive diode (PSD) (Pacific Sensor, DL16-7PCBA) located at a conjugate plane to the condenser back aperture. The PSD is connected to a field-programmable gate array (FPGA) multi-function acquisition card (National Instruments PCIe-7852R) and time series measurements of the nanowire position fluctuation with respect to the trapping centre are recorded at a sampling rate of 10kHz for a period of 10 seconds. The linearity of the PSD detector response with respect to the nanowire position in this orientation has been previously verified \cite{doi:10.1021/nl200720m}.

%
%


\begin{figure}[htb]
\centerline{\includegraphics[width= 0.85\linewidth]{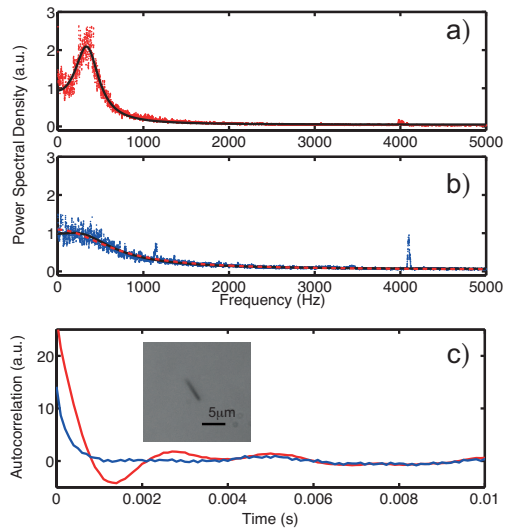}}
\caption{\label{fig:apsTOEfig2}(Color online) (a) Power spectral density of a single trapped nanowire exhibiting resonant oscillations (red) and the fitted spectrum (black) accounting for rotation / translational coupling and associated non-conservative effects. (b) Power spectrum of the same nanowire trapped close to the cover slip (blue) where the resonance significantly suppressed. The dotted (red) line is a Lorentzian fit and the solid (black) is the model. (c) Autocorrelation function of the nanowire motion at different heights showing resonant (red) and non-resonant behaviour (blue). The inset is a bright field image of the un-trapped nanowire.}
\end{figure}

The stochastic motion of a nanowire within an optical trap is normally described by a set of uncoupled Langevin equations in terms of translational ($x, y, z$) and angular coordinates ($\theta_{x}, \theta_{y}$). Each coordinate has an associated linear restoring torque/force, drag force and independent noise terms,  $\zeta_{i}(t)$ with zero mean and variance $\left<\zeta_{i}(t)\zeta_{i}(t+\tau)\right>= 2k_{b}T\gamma_{i}\eta(\tau)$.  The trapping experiments are performed in a low Reynolds Number aqueous environments and the associated drag coefficient, $\gamma$, is given by,

\begin{equation}\label{eq:drag}
\begin{split}
\gamma_{\perp}=\frac{4\pi\eta_{0}l}{\ln(l/2r)+\delta_{\perp}},  \\ \gamma_{\parallel}=\frac{2\pi\eta_{0}l}{\ln(l/2r)+\delta_{\parallel}}, \\
\gamma_{\theta}=\frac{\pi\eta_{0}l^{3}}{3(\ln(l/2r)+\delta_{\theta})}.
\end{split}
\end{equation}

Here $\gamma_{\perp}$ and $\gamma_{\parallel}$ are the drag coefficients perpendicular and parallel to the symmetry axis of the cylinder, and $\gamma_{\theta}$ is the rotational drag coefficient. $\eta_{0}$ is the viscosity of water  $\delta_{\perp}$, $\delta_{\parallel}$ and $\delta_{\theta}$ are correction factors accounting for the ends of the cylinder which depends on the ratio of the length ($l$) to diameter ($2r$) of the cylinder \cite{Tirado:1980aa, Tirado:1979aa, Broersma:1981aa}. Power spectrum analysis of this system yields a characteristic Lorentzian dependence for each degree of freedom where the corner frequency is given by the $f_{c}=K_{i}/2\pi\gamma_{i}$, where $K_{i}$ is the optical spring constant. 

A typical power spectrum of a trapped nanowire (blue) is presented in FIG~\ref{fig:apsTOEfig2}(b), together with the corresponding Lorentzian fit (dotted). Using the physical dimensions of 50 nm by 5 $\mu$m in length, (FIG~\ref{fig:apsTOEfig2}(c)[inset]), we estimated $K_{x}$ to be 2.9 pN$\mu$m$^{-1}$mW$^{-1}$ in this measurement, which is comparable to previously published results. Solving the Langevin equation for rotation yields a similar form of the power spectrum from which the restoring torque constant can be extracted. From previous work analysing autocorrelation data it is noted that the characteristic frequency for rotation is somewhat smaller compared with the translation motion and thus is not resolvable in these measurements \cite{Irrera:2011aa}.

When the trapping height above the cover slip is increased a broad resonant peak emerges in the power spectrum (red), shown in FIG~\ref{fig:apsTOEfig2}(a). The resonance peak persists only as the nanowire is trapped at a critical distance of about 50 $\mu$m from the cover slip, and reverts back to the normal response with decreased trapping height. The emergence of the resonance is also reflected in the autocorrelation function of the nanowire trajectory, where a fluctuation between positive and negative correlation can be seen, as shown in FIG~\ref{fig:apsTOEfig2}(c). The full height of the chamber is estimated to be in excess of 100$\mu m$ and so we exclude the possibility of direct interaction with the chamber walls. We note that changing the height of the trap will induce spherical aberrations and accentuate other aberrations already present within the trap. Increased aberrations will change the relative contributions from gradient and scattering forces for a given fixed input power, and we hypothesise that this is the origin of the height dependence of the resonance. 

In order to account for the emergence of this resonant peak we consider a more general form of the equations of motion which include cross-coupling terms. The dynamical equations in an over-damped environment can be written as a single 5-coordinate vector equation in terms of the generalized forces (neglecting rotation around the z-axis),

\begin{equation}\label{eq:coupled}
-\Gamma \dot{\vec{q}} + \vec{\zeta(t)} - K \vec{q} = \vec{0}.
\end{equation}

Here,  $\vec{q}=\left( x, \theta_{y}, y, \theta_{x}, z \right)$ are the generalised co-ordinates, $\Gamma$ is the diagonal hydrodynamic drag matrix, $\vec{\zeta(t)}$ is the Langevin force and $K$ is the trap stiffness matrix,

\begin{equation}\label{eq:stiffness}
K = \left(\begin{array}{ccccc}
                K_{x,x} & K_{x,\theta_{y}} & 0 & 0 & 0 \\ K_{\theta_{y},x} & K_{\theta_{y},\theta_{y}} & 0 & 0 & 0 \\ 0 & 0 & K_{y,y} & K_{y,\theta_{x}} & 0 \\ 0 & 0 & K_{\theta_{x},y} & K_{\theta_{x},\theta_{x}} & 0 \\ 0 & 0 & 0 & 0 & K_{z,z}
               \end{array}\right).
\end{equation}

Importantly, we only include off-diagonal terms in the stiffness matrix that couple between translation and rotation within a plane, i.e. a rotation in the xy plane ($\theta_{y}$) will induce a translation in x. This is justified as the non-conservative forces (radiation pressure) that are driving the coupling are principally directed along the z-axis and thus should induce rotation around the orthogonal axis. The above-mentioned approximations leads to two coupled equations of motion that retain the non-conservative effects, and that can be treated independently. An analysis of the two-coordinate stochastic equations leads to a power spectral density for each coupled coordinate (e.g. $i = x, \theta_{y}$) that has the form \cite{gardiner2004handbook}:

\begin{equation}\label{eq:psd}
S_{i}(\omega) = \frac{\eta_{i}\omega^{2}+\alpha_{i}}{2\pi((\omega^{2}-\epsilon)^{2}+(\mu\omega)^{2})},
\end{equation}

where $\mu$ and $\epsilon$ are the respective trace and determinant of the matrix $M = \Gamma^{-1}K$, and $\eta_{i}$ is associated with the noise strength. The term $\alpha_{x} = \eta_{x}m_{\theta_{y}\theta_{y}}^{2}+\eta_{\theta_{y}}m_{x\theta_{y}}^{2}$ is related to the cross-coupling matrix elements in $M$. In interpreting the solution in terms of the experimental system we note that if the coupling relates to the non-conservative scattering force, the off-axis components of the stiffness matrix should be non-symmetric, e.g. $K_{\theta_{y},x} \ne K_{x,\theta_{y}}$. The form of the power spectral density is similar to that of a damped harmonic oscillator, such as is observed in aerosol tweezing experiments \cite{PhysRevE.82.051123}. This is to be expected as both represent coupled first order differential equations; in the case of the underdamped oscillator coupling is between restoring and inertial terms, whilst in the present case it is between translation and rotation. A fit of the power spectrum presenting the resonant peak in FIG~\ref{fig:apsTOEfig2}(a) using Eq.~\ref{eq:psd} shows excellent agreement with the experimental data. In fitting the data we apply an additional offset to account for detector noise. Indeed a fit of the data taken at the lower trapping height FIG~\ref{fig:apsTOEfig2}(b) also reveals a strongly suppressed peak at a similar frequency. 



On inspecting the fitting parameters between of the different trapping heights we make the important empirical observation that a non-zero resonance peak is only observed under conditions where the one of the cross-coupling terms is negative. Such a scenario is ensured in this system as the trapping point is far from the centre of rotation - the action of rotation under Brownian motion will induces a translation at the trapping point which is in the opposite direction to the associated restoring force.

\begin{figure}[htb]
\centerline{\includegraphics[width= 0.85\linewidth]{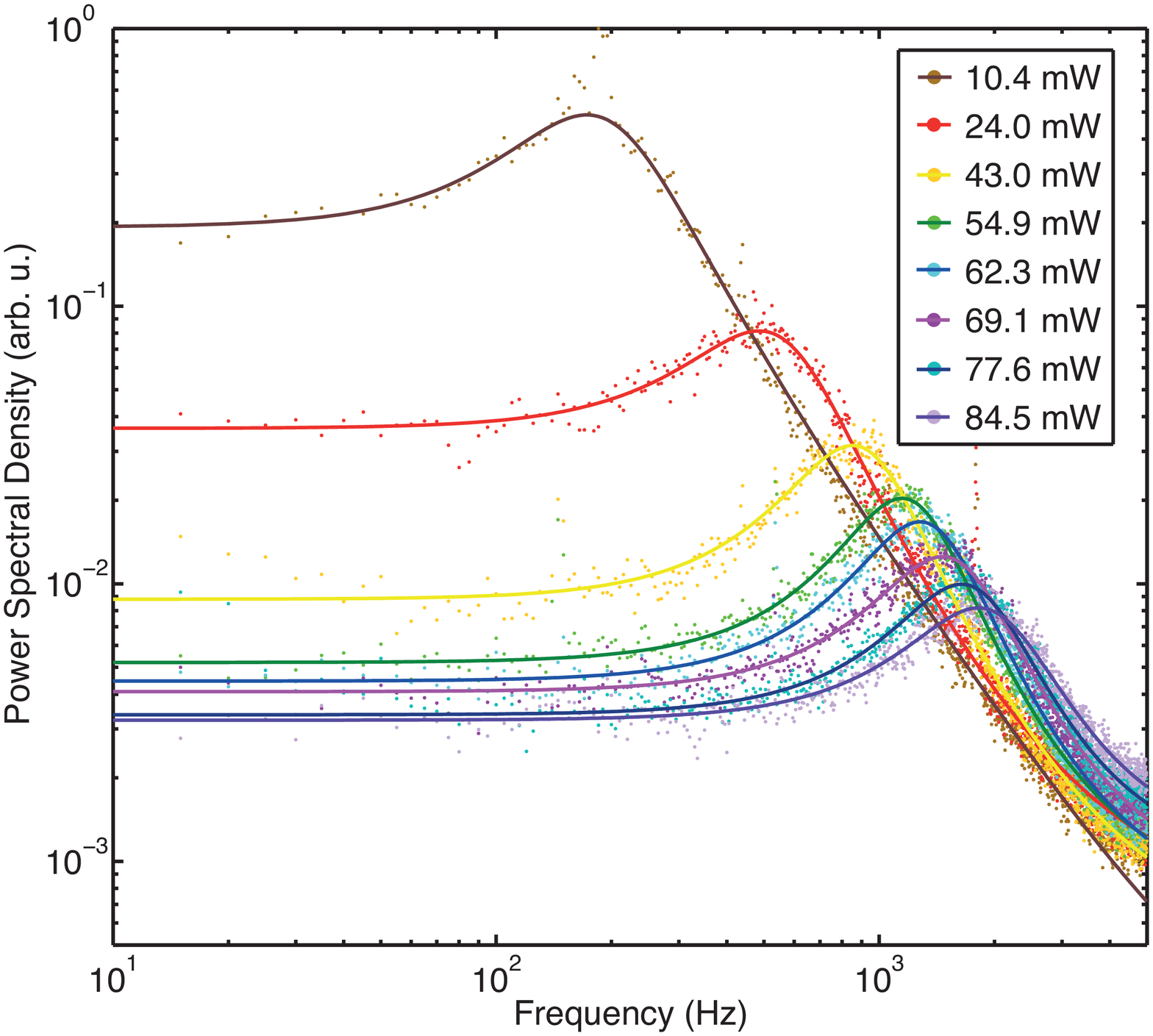}}
\caption{\label{fig:apsTOEfig3}(Color online) Power spectral density of a single nanowires held at a fixed trapping power measured for different trapping powers. The peak frequency shifts given by the $\sqrt{\epsilon}$ increases linearly with trapping power, whilst the Brownian fluctuations are strongly suppressed. The solid lines represent the fitted model which includes non-conservative cross-coupling between translation and rotation.}
\end{figure}

\begin{figure}[htb]
\centerline{\includegraphics[width= 0.9\linewidth]{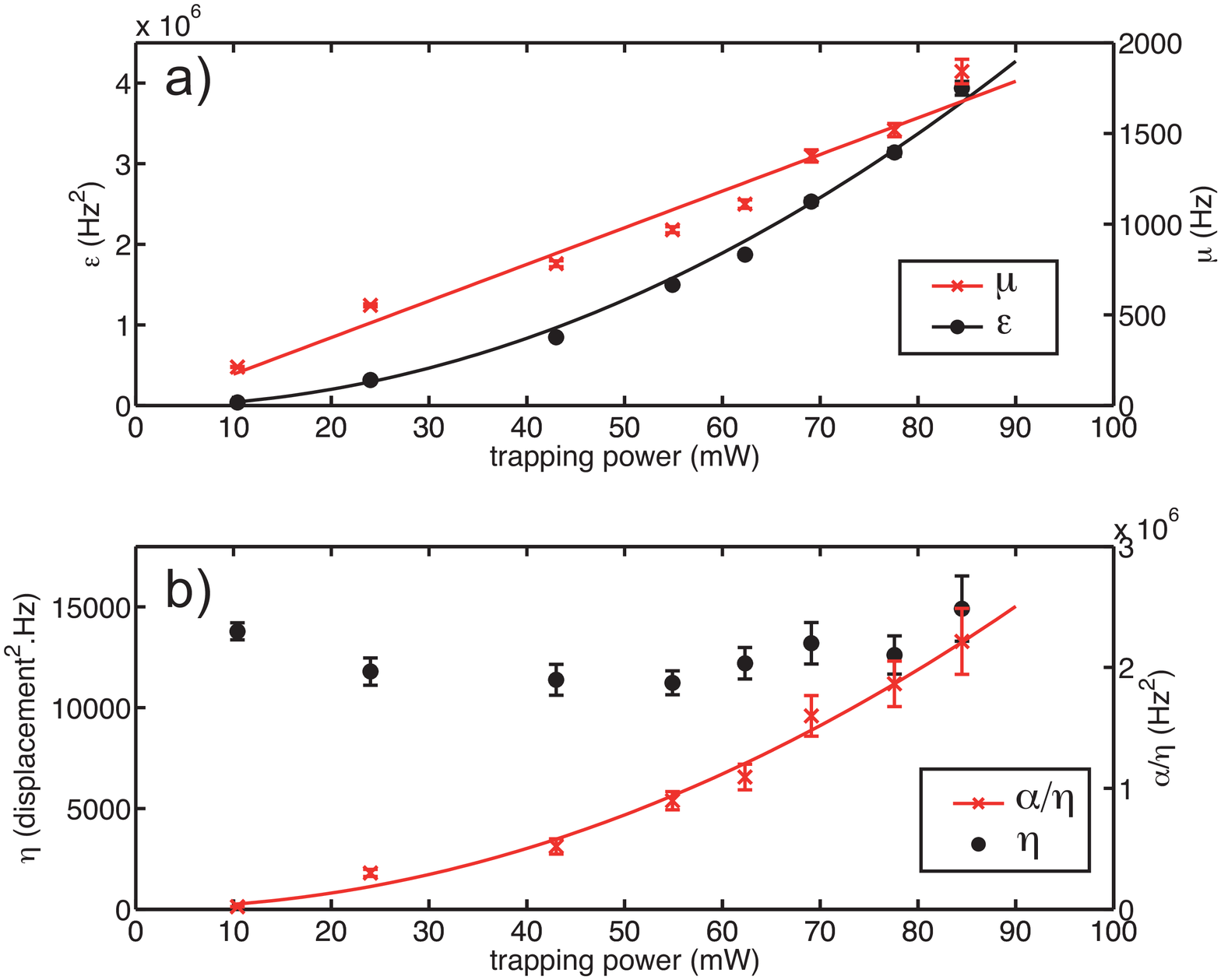}}
\caption{\label{fig:apsTOEfig4}(Color online) Fitting parameters of the power spectral density data presented in FIG~\ref{fig:apsTOEfig3}. (a) $\epsilon$ increase quadratically and $\mu$ linearly with trapping power, which corresponds to linear increase in both the peak frequency and width. (b) The noise strength, $\eta$, shows no clear dependence on trapping power indicating a constant temperature, whilst $\alpha$ increases quadratic with trapping power. All dependencies are consistent with a trap stiffness that is linearly proportional to the trapping intensity in the presence of asymmetric coupling.}
\end{figure}

To further investigate the dynamics of the resonance, the power spectrum of a trapped nanowire of length 4.6$\mu$m was recorded as a function of input trapping laser powers ranging from 10 mW to 90 mW. For low symmetry particles the trap stiffness scales linearly with power. The power dependent results are presented in FIG~\ref{fig:apsTOEfig3} along with the associated fits using Eq.~\ref{eq:psd}; again we note exceptional agreements between experimental results and the model. Qualitatively we see that the peak position of the resonance is blue shifted from 200 Hz to 1800 Hz with increasing trapping power together with a reduction of the amplitude of the fluctuations. The parameters associated with the fits are given in FIG~\ref{fig:apsTOEfig4}. We observe that $\epsilon$ increases quadratically with trapping power, whilst $\mu$ increases linearly. As the trap stiffness is proportional to trapping power and $\mu = tr(M)$ is proportional to the trap stiffness we should expect a linear increase in this parameter. Similarly, the as $\epsilon$ is the determinant of $M$ and should therefore be proportional to the square of the matrix elements, and hence proportional to the square of the power. $\eta$ is roughly constant at across all trapping powers, which indicates that there is no obvious heating across the different trapping powers.  Finally, we note that whilst the trapping power dictates the relative strengths of the trap stiffness (conservative) and radiation pressure (non-conservative), it should not modify the ratio of the two competing forces under conditions where the harmonic approximation for the trap stiffness is valid.
 
\begin{figure}[htb]
\centerline{\includegraphics[width= 0.9\linewidth]{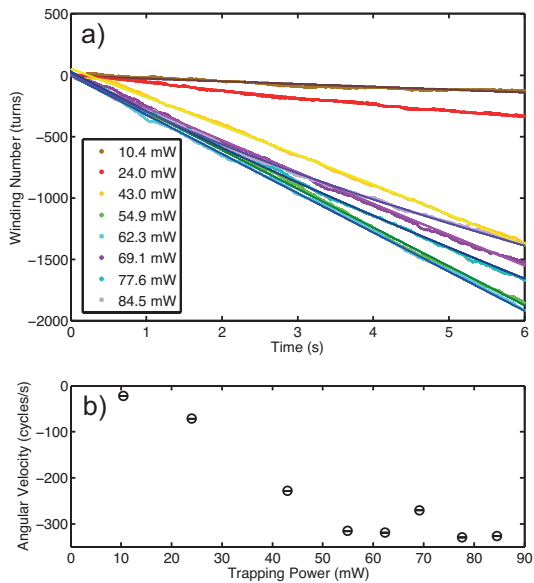}}
 \caption{\label{fig:apsTOEfig5}(Color online) (a) A winding number plot of the nanowire motion for different trapping powers. Small timescale fluctuations in the winding decorate a clear monatonic increase with time, indicating persistant cycling in one direction. (b) The rate of winding, given by the gradient of the winding plots, is observed to increase approximately trapping power for lower powers and plateau at higher powers. The winding also takes on a distinctly nonlinear trend at very high trapping frequencies.}
\end{figure}

In additional to the characteristic power spectrum described above, another important property of this system is its penchants to produce a continuous cycling of the position/orientation under steady state conditions. In order to discern the presence of cyclic motion a winding number analysis of the nanowire trajectories is performed. In our case the winding events are denoted when the nanowire trajectory moves around the centre of the trap in either a clockwise (negative) or anticlockwise (positive) motion. In an unbiased system the rates of cycling in either direction is balance leading to an average of zero winding. The results presented for different powers in FIG~\ref{fig:apsTOEfig5}(a), indicate that nanowires do indeed have a tendency to persistently cycle around a central point within the trap in a particular sense; short timescale fluctuations correspond to Brownian fluctuations in the trend. The winding behaviour is persistent over many seconds and is reproducible for each individually trapped wire. The sense of the cycling tends to be in a particular sense, suggesting an underlying bias within the trapping geometry, however winding in the opposite sense is also observed for some trapping events. We find that for different powers the rate of winding increases before plateauing at higher powers; this is clearly observed in a plot of the gradient of the winding number with trapping power, shown in FIG~\ref{fig:apsTOEfig5}(b). The winding property is dissipative in nature and is sustained only by the action of the non-conservative scattering forces. We may understand the process most intuitively by considering the following: when the nanowire is tilted with respect to the direction of the scattering force (along the z-axis) it will experience a greater pushing force. If the nanowire position lies below the focus it will be pushed through the focus to a position that is balanced by the axial restoring force. The additional lift force will only be alleviated when the tilt is reduced leading to a rotation back to the equilibrium position. The nanowire is then free to drop below the equilibrium position along the z-axis. 

In conclusion, we provide clear experimental and theoretical evidence for the influence of non-conservative coupling between translation and rotational modes in optically trapped nanowires. We see experimentally that this leads to a distinct resonance peak in the power spectrum and an accompanying winding of the particle trajectories. This model system provides exciting new opportunities to study complex dynamic behaviour involving non-conservative forces and may lead to exciting new insights into analogous biophysical systems.

\begin{acknowledgments}
The Australian National Fabrication Facility (ANFF) and Australian Research Council (ARC) are acknowledged for  supporting the nanowire growth facilities utilised in this work.
\end{acknowledgments}

\bibliography{WJTaps201X.bib}

\end{document}